\documentclass[twocolumn,showpacs,aps,superscriptaddress]{revtex4-2}
\usepackage{color}
\usepackage{soul}
\usepackage{lipsum}  
\usepackage{graphicx}
\usepackage{dcolumn}
\usepackage{bm}
\usepackage{hyperref}
\usepackage{amsmath}
\usepackage{mathtools}
\usepackage[export]{adjustbox}

\begin{document}

\preprint{APS/123-QED}

\title{Many-Body Colloidal Dynamics under Stochastic Resetting: Competing Effects of Particle Interactions on the Steady State Distribution}

\author{Ron Vatash}
\affiliation{The Raymond and Beverley School of Chemistry, Tel Aviv University, Tel Aviv 6997801, Israel.}
\author{Yael Roichman}
\affiliation{The Raymond and Beverley School of Chemistry, Tel Aviv University, Tel Aviv 6997801, Israel.}
\affiliation{The Raymond and Beverley School of Physics \& Astronomy, Tel Aviv University, Tel Aviv 6997801, Israel.}

\date{\today}

\begin{abstract}
 The random arrest of the diffusion of a single particle and its return to its origin has served as the paradigmatic example of a large variety of processes undergoing stochastic resetting. While the implications and applications of stochastic resetting for a single particle are well understood, less is known about resetting of many interacting particles. In this study, we experimentally and numerically investigate a system of six colloidal particles undergoing two types of stochastic resetting protocols: global resetting, where all particles are returned to their origin simultaneously, and local resetting, where particles are reset one at a time. Our particles interact mainly through hard-core repulsion and hydrodynamic flows. We find that the most substantial effect of interparticle interactions is observed for local resetting, specifically when particles are physically dragged to the origin. In this case, hard-core repulsion broadens the steady-state distribution, while hydrodynamic interactions significantly narrow the distribution. The combination results in a steady-state distribution that is wider compared to that of a single particle system both for global and local resetting protocols.  
 
\end{abstract}

\maketitle

Stochastic resetting (SR), the abrupt restarting of a process at random time intervals, drives the system toward a steady state. When applied to a diffusing particle, it leads to the emergence of a stationary density profile \cite{evans11, Evans20, ofir20}. SR induced steady-states have been used to model a wide range of physical and natural phenomena \cite{Evans20, gupta2022, Nagar23ManyBodyReset}, including the height distribution of fluctuating interfaces \cite{Gupta2014, Gupta2016} and the mechanisms governing cell division \cite{Genthon2022}.

For exponentially distributed resetting times at a constant rate, the steady-state density follows a Laplacian form, determined by the ratio of the resetting rate to the diffusion coefficient \cite{evans11}. This density profile, however, changes when the resetting time distribution is altered \cite{Roldan2017} or when partial resets are introduced \cite{OfirPartialRestart22, MetzlerSoftResettingHarmonicPotential22, Metzler23PartialRestart}. Further complexity arises when memory effects or ballistic motion are incorporated into SR processes \cite{BoyerResetwithMemoryPRL14, Majumdar24_PRE_ActiveParticleWithMemory}. The ability to control and engineer steady-state profiles through SR opens up potential applications, such as accelerating state transitions \cite{Remi24}.

While understanding SR in single-particle systems provides a foundation, its impact on multi-particle systems is crucial for addressing real-world conditions. Numerous phenomena, including reaction kinetics \cite{shlomi15michaelismen, Pal_2022Inspection}, foraging  \cite{Amy24, Pal20Foraging, Ofek24Sokoban}, and queuing \cite{Ofek23Queues}, inherently involve interacting processes.
Pioneering studies on many-body systems under SR \cite{MironReuveni21, Nagar23ManyBodyReset, biroli23_Majumdar} define global and local resetting protocols. The former refers to a process in which the entire system is reset at each resetting event. In contrast, only one of the particles is returned to the origin at each resetting event in the latter. These protocols can lead to significant differences in the emergent steady state. For instance, in a system of particles diffusing on a ring, \cite{MironReuveni21}, these two protocols resulted in qualitative differences in the evolving steady-state distribution of the particle distribution. Interestingly, resetting only a small subset of coupled oscillators can lead to full system synchronization \cite{Majumder2024}. 
Analytical calculations of stochastic resetting (SR) properties in many-body systems are challenging due to their complexity. Consequently, experimental studies are crucial for understanding how different interparticle interactions affect systems under SR. 
\begin{figure}[t]
    \begin{center}
    \includegraphics[scale = 0.35]{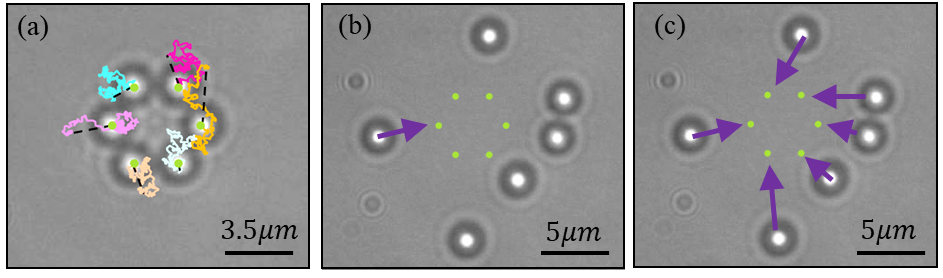}
    \caption{Colloidal particles under stochastic resetting to a hexagonal configuration using automated holographic optical tweezers. (a) Particles trapped in their initial arrangement with trajectory segments between resetting events superimposed. Green dots indicate the positions of the optical traps, and black dashed lines indicate the return routes. Two resetting protocols are illustrated: (b) Local resetting and (c) Global resetting protocols. Purple arrows indicate particle movement during a reset.}
    \label{fig:ExpDesign}
    \end{center}
\end{figure}

In this letter, we present an experimental and numerical investigation into the impact of realistic interparticle interactions on steady-state distributions arising from stochastic resetting. We employ optical trapping techniques to experimentally reset colloidal particles suspended in water. Initially, we measure the free propagator of a particle within a many-body system and quantify how interparticle interactions alter its shape relative to the single-particle case. Subsequently, we investigate the effect of these interactions on the steady-state distribution under various resetting protocols. Through numerical analysis, we differentiate the contributions of distinct interaction types and uncover competing influences on the steady-state distribution evolution. Finally, we demonstrate that the specific impact of each interaction strongly depends on the chosen resetting protocol.


Specifically, our experiments are conducted using holographic optical tweezers (HOTs) \cite{Dufresne_Grier01,Polin05optimizedTraps} to manipulate the positions of colloidal particles (silica, diameter $d=1.5\pm0.08$~$\mu m$) suspended slightly above a glass sample floor. The particles primarily interact through partially suppressed hydrodynamic interactions  \cite{Grier06Hydro, 
 SokolovRoichmanHydro11, Svetlizky2021} and near hard-core repulsion \cite{RoelPNAS_06_Directmeasurment, AliceRoelPRL17_MeltingColloidal, KapfenbergerAdar_HolographicImagingOpticExpress13}. At the start of each experiment, particles are optically trapped in a hexagonal configuration around the origin, with an inter-trap distance of $R_0=2.5\pm0.03~\mu$m. The particles are then released and are allowed to diffuse freely (Fig.~\ref{fig:ExpDesign}(a)).  

An automated computer protocol manages the resetting process, beginning by determining the time for the next reset event. When this time arrives, the system captures an image of the sample. Using conventional image analysis \cite{CROCKER1996VideoAnalysis}, the particle's location is determined. The protocol then projects optical traps onto the particles designated for reset, manipulating them to guide each selected particle back to its original position in the hexagonal configuration. After successfully resetting all designated particles, the optical traps are deactivated, releasing the particles. The countdown for the next reset event starts immediately, with the interval between events randomly determined from an exponential distribution with a mean resetting rate of $r$. To enable a direct comparison of local and global returns (Fig.~\ref{fig:ExpDesign}(b,c)), both were conducted under an approximately fixed-duration resetting protocol. i.e., the return phase duration is fixed and is independent of the return distance. 

To complement our experiments and distinguish between hydrodynamic interactions and hard-core repulsion, we conduct Stokesian Dynamics simulations \cite{Brady88, SokolovRoichmanHydro11, HarelYael14}. Collisions are modeled using a repulsive Weeks-Chandler-Anderson (WCA) pair interaction \cite{WCA}, while hydrodynamic interactions are captured through the many-body Rotne-Prager mobility tensor \cite{rotneprager}. Temperature effects are introduced via a Gaussian-distributed random force that adheres to the fluctuation-dissipation relation, utilizing the same mobility tensor.

For a single particle under SR and many-particle systems under global resetting, it is common to treat separately the free diffusion and the return phases of the process \cite{ofir20, BordovaSokolov20PRE_noninstantaneuosreseting}. In such cases, the contribution of the free diffusion phase to the steady-state distribution can be calculated using the renewal equation,
\begin{equation}
    \label{renewaleq}
    p(x, t| x_0) = e^{-r t} C(x, t|x_0) + r \int_{0}^t d \tau e^{-r\tau} C(x, \tau|x_0), 
\end{equation}
where $r$ is the resetting rate and \(C(x,t|x_0)\) is the free diffusion propagator, i.e., the reset-free probability density function \cite{evans11}. We note that this separation is impossible for local resetting in which a particle is returned while other particles continue diffusing. 

In the case of normal diffusion, where \(C(x, t)\) follows a Gaussian distribution, the well-known steady-state solution is obtained \cite{evans11},
\begin{equation}
    \rho(x|x_0) = \frac{\alpha_0}{2} \exp\left(-\alpha_0 |x - x_0| \right),
    \label{eq:SS_x_SingleParticle}
\end{equation}
where, \(\alpha_0 = \sqrt{r/D}\) represents an inverse length scale corresponding to the characteristic distance a particle diffuses between reset events.

\emph{Effect of particle interaction on the diffusion propagator} - To evaluate the impact of particle interactions on the free diffusion propagator, we begin by tracking the motion of a single particle released from an optical trap and extracting its trajectory using conventional image analysis \cite{CROCKER1996VideoAnalysis}. This process is repeated 15,000 times, with each release lasting two minutes. Next, we simultaneously released six particles from their respective traps and tracked their motion for the same duration, repeating the experiment 500 times. 
\begin{figure}[t]
    \centering
    \includegraphics[width=0.48\textwidth]{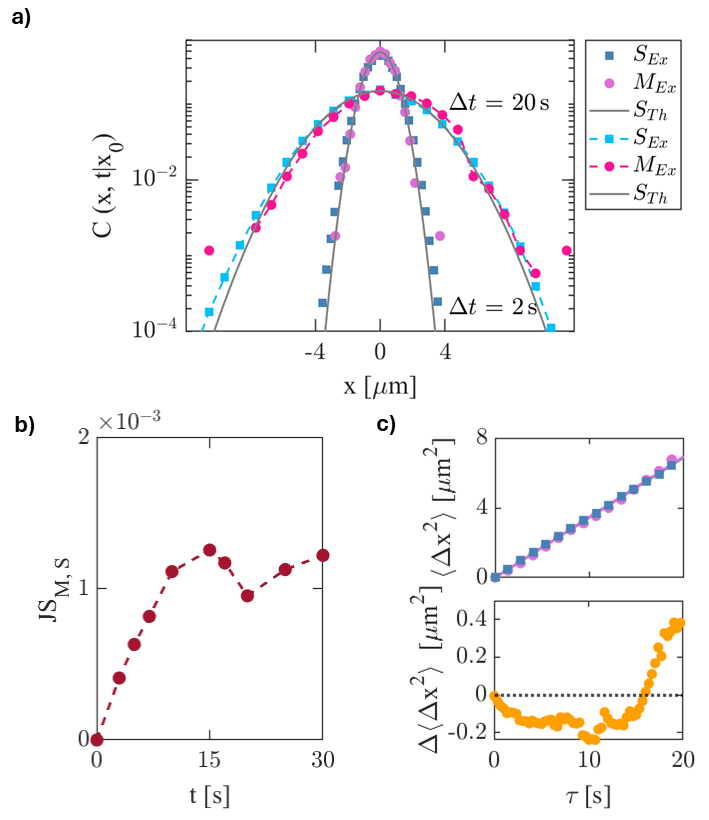}
    \caption{a) Reset-Free Propagators $C(x,t|x_0)$ of a single and six particle systems at $t = 2\thinspace{s}$ and $t=20\thinspace{s}$.  Single (Six) particle experimental measurements are presented by blue squares (pink circles), and the normal diffusion Gaussian propagator with $D = 0.18\thinspace{\mu m^2/s}$ is presented by a gray line. b) The JS distance between a single and six particle propagators as a function of time. c) Ensemble average mean square displacement (MSD) as a function of lag time  $\tau$ of single (blue) and six-particle (pink) systems (top panel) and the difference between them (lower panel).}
    \label{fig:resetfreepropSinglevsMany}
\end{figure}
The experimentally measured free propagator of a single particle is in excellent agreement with the theoretical prediction at both short times ($t=2$\thinspace{}s) and longer times ($t=20$\thinspace{}s), as shown in Fig.\ref{fig:resetfreepropSinglevsMany}(a). When comparing the free propagator of the six-particle system to that of a single particle, we find that they coincide at short times, but at longer times, a slight shift of the probability density toward larger distances emerges (Fig.\ref{fig:resetfreepropSinglevsMany}(a)). 

We quantify how the single-particle and six-particle propagators, $C_s$ and $C_m$, deviate over time using the Jensen–Shannon divergence (Fig.~\ref{fig:resetfreepropSinglevsMany}(b)). At short times, the divergence remains small since the particles have not yet interacted significantly via hard-core repulsion. As time progresses, the divergence increases, reaching a peak at intermediate times when particles come into close proximity, and their interactions become more pronounced, leading to more significant differences between the single-particle and many-body distributions. We expect these differences to emerge on a timescale related to the diffusion time required for particles to traverse their initial separation, which aligns with our findings. At longer times, as the particles disperse and their interactions diminish, the divergence stabilizes due to the reduced influence of hard-core repulsion.

A more detailed comparison of the free propagator’s evolution in both systems can be made by examining the temporal evolution of the mean square displacement. For the single-particle system, it follows the expected linear dependence on lag time $\tau$ (Fig. \ref{fig:resetfreepropSinglevsMany}(c)).  In contrast, the mean square displacement of the six-particle system does not follow a strictly linear dependence on lag time. At short times, it grows more slowly than in the single-particle case, indicating a reduced effective diffusion constant,  $D_m = 0.16 \pm 0.03~\mu m^2/s$ compared to $D_s = 0.18 \pm 0.03~\mu m^2/s$. In this regime, before the particles diffuse far enough to collide, they remain relatively close to each other and primarily interact through hydrodynamic interactions. The higher colloidal density results in a higher effective viscosity \cite{Sonn-Segev2015}, reducing the diffusion coefficient. At longer timescales, collisions between colloidal particles facilitate faster spreading, resulting in an increased mean square displacement, as shown in Fig.~\ref{fig:resetfreepropSinglevsMany}(c).


\emph{Steady state under stochastic resetting} - We determine the steady-state distribution of particle positions under stochastic resetting by leveraging reset-free diffusion data. This approach allows us to analyze a wide range of resetting conditions using the same set of experiments \cite{vatash2025numerical}. 
Our analysis follows a systematic procedure that applies to single and six-particle systems under global resetting. First, we generate a sequence of random resetting times, $\{t_1, t_2, ...\}$, drawn from an exponential distribution with a mean resetting rate $r$. Next, each recorded reset-free trajectory is truncated at its corresponding reset time $t_i$. These modified trajectories are then seamlessly concatenated, creating an extended sequence representing a system undergoing instantaneous global resetting (teleportation). Finally, we use this constructed sequence to compute the steady-state distribution under global and instantaneous resetting protocol.

Alternatively, the steady-state distribution under resetting can be evaluated from the measured free propagator by numerically integrating Eq.~\ref{renewaleq} and taking $t\rightarrow t_{\text{ss}}$ \cite{vatash2025numerical}, 
\begin{equation}
    \label{eq:RenewalSumForm}
    \rho(x|x_0) \approx r \sum_{i=1}^{t_{\text{ss}}} \Delta t \Psi( t_i) C(x, t_i| x_0),
\end{equation} 
where $t_{\text{ss}}\approx 60$\thinspace{}s is the time it takes the system to reach steady-state in our system, and $\Psi(t)=  \int_t^\infty r e^{-r t}dt$ is the probability of a process to survive without resetting until time $t$.

\begin{figure}[h]
    \centering
    \includegraphics[scale = 0.5]{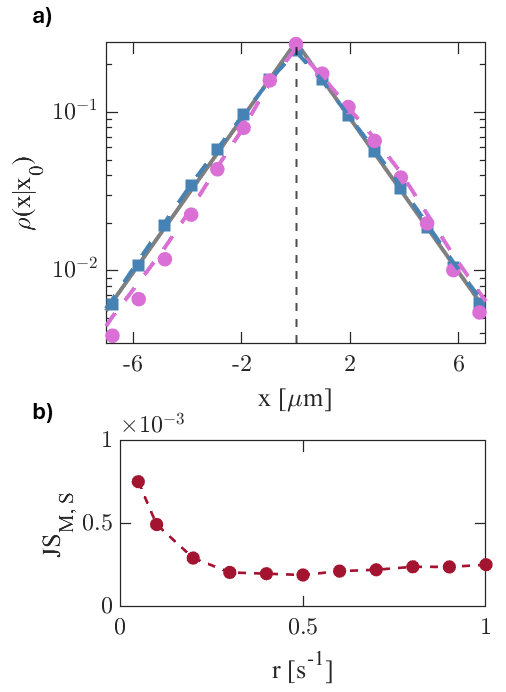}
    \caption{(a) Steady-State Distributions comparing single to six particle systems under a resetting rate of $r=0.05$\thinspace{s$^{-1}$}. Experimental measurements for single (blue squares) and six (pink circles) particles agree well with the renewal prediction (blue and pink dashed lines, respectively). For the single particle system, the theoretical prediction according to Eq.\ref{eq:SS_x_SingleParticle} is plotted in a solid gray line. (b)  }    
    \label{fig:steadystaeManyvsSingleExpVsTheoVsRenewal}
\end{figure}

For each particle, we compute the steady-state distribution under resetting along the line connecting the center of the hexagon to its initial position. We obtain the final distribution for the six-particle system by averaging over all individual particle distributions. 
In Fig.~\ref{fig:steadystaeManyvsSingleExpVsTheoVsRenewal}(a), we compare the steady-state distributions of the single- and six-particle systems. For the single particle system, the distributions obtained from experiments (symbols), the numerical renewal equation (dashed lines), and Eq.~\ref{eq:SS_x_SingleParticle} (solid line) show strong agreement. Comparing the single-particle (blue) and six-particle (pink) steady-state distributions, we observe a transition to an asymmetric steady-state distribution in the six-particle system. This observation aligns with our earlier findings on interaction effects in the free propagator, where collisions enhance spreading at large distances and intermediate times but constrain motion at smaller radii.

The JS distance between the single and six particle systems decreases with increasing resetting rates, reaching an approximately constant low value at $r=0.5$\thinspace{s$^{-1}$} (Fig.~\ref{fig:steadystaeManyvsSingleExpVsTheoVsRenewal}(b)). The results clearly indicate that as the resetting rate decreases, the difference between the distributions becomes more pronounced, corresponding to the increased difference in the free propagator over a long period of time. The particles have more opportunities to interact for the lower measured resetting rates, resulting in a more significant difference between the steady-state distribution of the single-particle system and the many-body system.

\begin{figure}[t]
    \centering
    \includegraphics[scale=0.5]{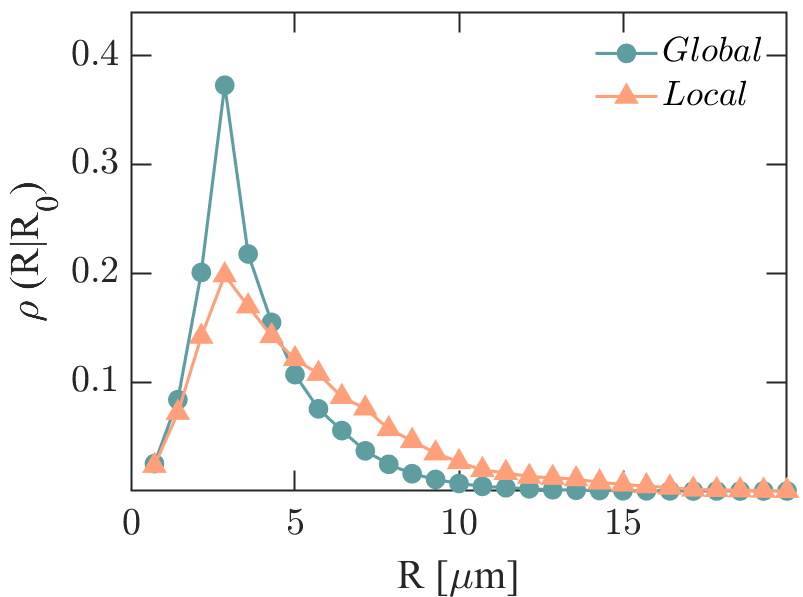}
    \caption{Experimentally measured radial steady-state distribution for the local resetting (orange triangles) and global resetting (blue circles) protocols. In both cases, the resetting rate is \( r = 0.05\thinspace{s^{-1}} \), with an average return time of \( t_{\text{rtn}} = 6.20\thinspace{s} \).}
    \label{fig:GlobalVsLocalExperiments}
\end{figure}

Although particle interactions alter both the free propagator and the steady-state distribution, their overall impact under global resetting remains limited (Fig.~\ref{fig:steadystaeManyvsSingleExpVsTheoVsRenewal}). We attribute this to experimental conditions that minimize hydrodynamic interactions and to the idealized protocol of instantaneous returns upon resetting. A more substantial influence of particle interactions is expected if returns involve physically dragging the particle back to the origin in a local resetting scheme, during which other particles continue their diffusive motion. 

We implemented a local resetting protocol with constant-time returns, in which particles are actively dragged back to the origin, each at a time,  using optical tweezers. As a result, particles can interact with other particles along their return paths. To enable direct comparison with global resetting, we set the local resetting rate to be six times faster than the global resetting rate. In experiments, since returns are performed by a moving laser trap, a diffusing particle that comes close to a resetting particle can also get pulled in by the same laser beam; consequently, both particles end up returning together. We find that approximately $20\%$ of the resetting events, out of 975 recorded, involve multi-particle resets. 


In Fig \ref{fig:GlobalVsLocalExperiments} we plot the steady-state distribution in the radial direction along the lines connecting the origin to each vertex of the hexagon, averaging over all six directions. The effect of particle interactions is immediately evident when comparing local resetting (orange) to global resetting (blue). In the case of local resetting, the probability density function is significantly broader. We attribute this to the specific resetting protocol we use.

The resetting protocol we implement starts the clock for the next resetting event only after the returning particle has reached the origin. Consequently, for local resetting, when one particle resets, the remaining five continue diffusing, whereas, during global resetting, each event returns all six particles at once. This difference leads to a longer net diffusing time per particle under local resetting, thereby broadening the steady-state distributions relative to global resetting.


We performed two sets of computer simulations to decouple the contributions of hard-core repulsion and hydrodynamic interactions to the six-particle steady-state distribution. The first included only hard-core repulsion, while the second employed Stokesian dynamics that incorporates hydrodynamic interactions in an idealized, infinite fluid. We also implemented the simulations in such a way that only one particle interacts with the resetting laser at a time, allowing us to avoid spurious multi-particle returns. A further distinction between simulations and experiments are the hydrodynamic interactions: in the experiments, an absorbing boundary alters fluid flow around the particles and weakens their coupling, whereas in the simulations, the particles experience stronger, bulk-like hydrodynamic interactions. 

The simulation resetting protocols mirrored those of the experiments. A total of 10,000 return events had been recorded. Following our experimental design, we compare global and local protocols while maintaining a constant per-particle resetting rate, albeit with different free diffusing times per particle. 

\begin{figure}[h]
    \centering
    \includegraphics[scale = 0.4]{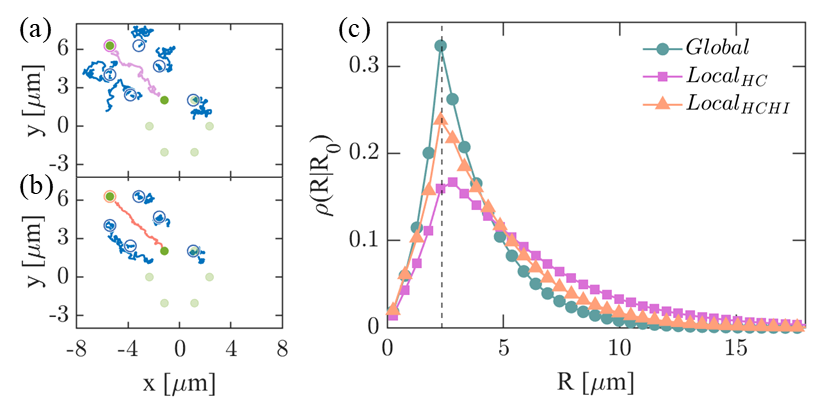}
    \caption{Global vs. Local returns; a) and b) typical trajectories of the resetting particle (pink and orange), and the other five particles (blue) during a single reset event in without and with hydrodynamic interactions, respectively. c) Radial steady-state distribution form simulated data of different resetting protocols and particle interactions: global resetting with $r=0.3$ with hydrodynamic interactions (blue circles), local resetting with $r=0.05$ without hydrodynamic interactions (pink squares), and with hydrodynamic interactions (orange triangles). All simulations involved a constant return time protocol (3 seconds). The black dashed line represents the initial and resetting position.}
    \label{fig:globalvslocalsims}
\end{figure}
Figure~\ref{fig:globalvslocalsims}(a) and (b) illustrate simulated trajectories of a particle suspension where one particle is actively reset to the nearest trapping position, both with and without hydrodynamic interactions. Light green circles show the reset positions, the particle trajectories are traced in blue, and the particle undergoing resetting is highlighted in green, with its trajectory in pink (without hydrodynamic interactions) or orange (with hydrodynamic interactions). A striking difference emerges between the two scenarios: in the absence of hydrodynamic interactions, the resetting particle merely pushes away any particle it collides with, whereas, in their presence, the fluid flow created by its motion pulls nearby particles along in the same direction.

In Fig.~\ref{fig:globalvslocalsims}(c), we plot the steady-state distribution in the radial direction along the line connecting the origin with each hexagon vertex, averaged over all six directions. Local resetting, both with (orange triangles) and without (pink squares) hydrodynamic interactions, yields a wider distribution than that of global resetting, reflecting the impact of collisions during return (as seen in Fig.~\ref{fig:globalvslocalsims}a) that do not occur in global resetting. Notably, the resetting rate per particle is  $r=0.05\thinspace{s^{-1}}$ in all simulations, meaning the total local resetting rate is effectively six times faster. Furthermore, in the absence of hydrodynamic interactions, the steady-state distribution broadens more, indicating that hydrodynamic interactions partially mitigate the collisional effect by narrowing the distribution.

An intuitive explanation for this narrowing effect arises from the distance dependence of hydrodynamic interactions. In Fig.~\ref{fig:globalvslocalsims}(b), the resetting particle generates a flow toward the nearest hexagon vertex, but the key factor is how strongly neighboring particles are pushed inward versus outward. Particles nearer to the resetting particle experience a stronger inward push, whereas those on the opposite side of the hexagon feel a weaker outward push, simply by virtue of being farther from the flow source. When averaged over many return events, this asymmetry narrows the steady-state distribution.

\emph{Conclusions and discussion}

In this paper, we used holographic optical tweezers to implement different stochastic resetting protocols on a colloidal suspension. Our approach combines experiments with computer simulations to examine how inter-particle interactions shape the emergent steady-state distribution. We observe that collisions and hydrodynamic effects play a key role, particularly during the resetting process, underscoring the importance of interactions in determining the steady-state distribution.

For global instantaneous resetting, where particles are returned to the origin simultaneously, the measured steady-state distribution can still be predicted from the many-body propagator using the renewal approach, as in single-particle resetting. In our implementation, we reset particles by optically trapping and translating them to the origin in a controlled manner. This procedure largely suppresses interactions during the return phase, allowing the system to remain accurately described by the renewal framework in combination with known particle trajectories. The resulting agreement between theory and experiments suggests that renewal approaches can be valid even in many-particle systems with collisions and hydrodynamic coupling.

By contrast, local resetting does not simply reset all particles at once; only a single particle is returned while others continue diffusing. This blends the return phase and the exploration phase, so the free propagator alone is no longer sufficient to predict the steady-state distribution. In our experiments, we also observe unintended partial returns, where non-targeted particles are inadvertently dragged partway—or even entirely—back to the origin during a local resetting event. These additional interactions further affect the steady-state distribution, introducing a partial-resetting component not captured by the single-particle propagator. Overall, this interplay underscores the pronounced influence of many-body effects in shaping the final distribution when local resetting is used.

Overall, our results demonstrate that the resetting protocol, both in terms of sequence and timing, plays a pivotal role in how interactions shape the steady-state distribution, particularly when the return and exploration phases are mixed. Moreover, different types of interactions can compete, either broadening or narrowing the distribution. Finally, given that stochastic resetting is increasingly employed to accelerate search processes, our findings offer valuable insights into designing many-body search strategies in the presence of inter-particle interactions.

%

\end{document}